\documentclass[fleqn,useAMS,usenatbib]{mn2e}
\usepackage{epsfig,subfigure,amsmath}
\usepackage{color}

\definecolor{purple}{rgb}{1,0,1}

\newcommand{\lcdm}{$\Lambda$CDM}
\newcommand{\hmpc}{$h^{-1}$Mpc}
\newcommand{\gmpc}{$h^{-1}$Gpc}

\newcommand{\beq}{\begin{equation}}
\newcommand{\eeq}{\end{equation}}

\bibpunct[; ]{(}{)}{;}{a}{}{,}

\title{Using cosmic voids to distinguish f(R) gravity in future galaxy surveys}

\author[Zivick et al.]
{
\parbox{\textwidth}{
Paul Zivick$^{1,2}$,
P.~M. Sutter$^{2,3,4,5,6}$,
Benjamin D. Wandelt$^{5,6,7,8}$, 
Baojiu Li$^{9}$,
and Tsz Yan Lam$^{10}$
}
\vspace{0.4cm}\\
\parbox[c]{\textwidth}{
$^{1}$ Department of Astronomy, Ohio State University, Columbus, OH 43210 \\
$^{2}$ Center for Cosmology and Astro-Particle Physics, Ohio State University, Columbus, OH 43210\\
$^{3}$ INFN - National Institute for Nuclear Physics, via Valerio 2, I-34127 Trieste, Italy \\
$^{4}$ INAF - Osservatorio Astronomico di Trieste, via Tiepolo 11, 1-34143 Trieste, Italy \\
$^{5}$ Sorbonne Universit\'{e}s, UPMC Univ Paris 06, UMR7095, Institut d'Astrophysique de Paris, F-75014, Paris, France \\
$^{6}$ CNRS, UMR7095, Institut d'Astrophysique de Paris, F-75014, Paris, France \\
$^{7}$ Department of Physics, University of Illinois at Urbana-Champaign, Urbana, IL 61801, USA \\
$^{8}$ Department of Astronomy, University of Illinois at Urbana-Champaign, Urbana, IL 61801, USA \\
$^{9}$ Institute for Computational Cosmology, Department of Physics, University of Durham, South Road, Durham DH1 3LE, UK \\
$^{10}$ Max Planck Institute for Astrophysics, Karl-Schwarzschild-Str. 1, 85748 Garching, Germany
}}

\begin{document}

\maketitle

\label{firstpage}

\begin{abstract}
We use properties of void populations identified in $N$-body
simulations to forecast the ability of upcoming galaxy surveys 
to differentiate models of f(R) gravity from \lcdm~cosmology. 
We analyze multiple simulation realizations, which were designed 
to mimic the expected number densities, volumes, and redshifts  
of the upcoming Euclid satellite and a lower-redshift ground-based 
counterpart survey, 
using the public {\tt VIDE} toolkit. 
We examine void abundances, ellipicities, radial density profiles,
and radial velocity profiles at redshifts 1.0 and 0.43. 
We find that
stronger f(R) coupling strengths eliminates small voids and 
produces voids up to $\sim 20\%$ larger
in radius, leading to a significant tilt in the void number
function. Additionally, under the influence of modified gravity,
voids at all scales tend to be measurably emptier with correspondingly higher
compensation walls. The velocity profiles reflect this, showing
increased outflows inside voids and increased inflows outside voids.
Using the
void number function as an example, we forecast that future surveys 
can constrain the modified gravity coupling strength to $\sim 3 \times 10^{-5}$ using voids.
\end{abstract}


% -----------------------------------------------------------------------------
% -----------------------------------------------------------------------------
\section{Introduction}

While current cosmological tests show that the inflation plus cold dark matter (\lcdm) paradigm can successfully describe the observational properties of the universe~\citep[e.g.,][]{Reid2012,Planck2013}, it does not explain the nature of dark matter or dark energy. In order to explain observational results while avoiding the inclusion of dark energy, alternative theories have been proposed, including the modification of gravity. The motivation stems from the fact that gravity acts as the main interaction at large scales and accordingly has shaped the evolution of the universe, so perhaps there is a more accurate description of gravity that could account for questions left unanswered by \lcdm. While there are many different proposed modifications~\citep[e.g.,][]{Dvali:2000,Maartens2004}), this paper will focus on a single example from the \textit{f(R)} class of models, which contain relatively simple modifications to GR and act as a useful tool to build a better understanding of the potential effects of modified gravity. The specifics of the model used will be discussed in the following section. 

A key feature of the \textit{f(R)} model is the presence of the chameleon mechanism~\citep{Khoury:2004}. One problem frequently encountered in proposed modified gravity models is the necessity to build them so as to pass current solar system tests. However, any model that strengthens gravity will struggle to pass in overdense regions such as the solar system. That is where the chameleon mechanism comes into effect, by coupling the gravity modifications to the local density, suppressing the effects of modified gravity in high density regions while allowing it to be unscreened in underdense regions. Past studies have tested \textit{f(R)} models for different observational signatures, from the degree of curvature in superclusters~\citep{Shim2014}, which found that superclusters tended to be straighter in \lcdm, to galaxy population statistics~\citep{Fontanot2013}, which found that there was no significant difference between \lcdm~and \textit{f(R)}, to the ISW effect on the power spectrum~\citep{Cai2014}, which found potentially detectable differences between the models in both the linear and nonlinear regime.

Perhaps a more natural choice to test \textit{f(R)} gravity would be to focus on the regions where it would be unscreened. These underdense regions, commonly referred to as voids, could then provide a means of distinguishing between \lcdm~cosmology and modified gravity models. Already these voids have been used as a potential diagnostic for examining coupled dark energy-dark matter models~\citep{Sutter2014d}, and have been used in applications such as 
weak anti-lensing~\citep{Melchior2014,Clampitt2014}, the Alcock-Paczynski 
test~\citep{LavauxGuilhem2011, Sutter2012b, Sutter2014c}, and the integrated Sachs-Wolfe effect~\citep{Planck2013b, Ilic2013}, demonstrating the usefulness of voids as cosmological probes. With regard to modified gravity and voids, some efforts have been made already to explore this avenue~\citep{Li:2012, Clampitt2013}, and initial results provide hints that voids may indeed be a viable testbed.

Currently available void catalogs~\citep{Pan2011,Sutter2012a,Sutter2013c} from the SDSS galaxy surveys~\citep{SDSS:2009,Ahn2012} have enabled for the first time direct comparisons of predicted void characteristics to actual survey data for low redshifts. And current void-based studies of modified gravity have  analyzed both redshift 0 conditions as well as higher redshift conditions through the use of a spherical underdensity void finding algorithm~\citep{Cai2014b}. However, upcoming surveys such as Euclid~\citep{Euclid:2011} and WFIRST~\citep{Spergel2013} will target higher redshifts, unlocking a tremendous number of voids~\citep{Pisani2014b}. It is thus necessary to provide a proper forecast of the ability of voids found through the watershed transform to distinguish modified gravity models in this new observing regime.

In this paper we present an assessment of the impact modified $f(R)$ 
gravity models on various void statistics, such as number functions, 
ellipticities, and radial density profiles. We focus on simulations 
modeling the higher redshifts, large volumes, and sparse densities
 comparable to upcoming 
surveys to look for observationally-viable distinguishing characteristics 
of the voids produced by the different models.

In the following section, we briefly discuss the simulations analyzed 
and the toolkit used for finding voids. In Section~\ref{sec:effects} 
we discuss the effects of modified gravity on void characteristics 
and conclude in Section~\ref{sec:conclusions} with the implications 
for future surveys and potential options for more refined forecasts.

% -----------------------------------------------------------------------------
% -----------------------------------------------------------------------------
\section{Simulations \& Void Finding}
\label{sec:approach}

The particular class of modified gravity theories analyzed in this paper, denoted by \textit{f(R)}, is marked by the generalization of the Ricci scalar \textit{R} in the Einstein-Hilbert action. When combined with the chameleon mechanism, the structure formation equations become
\begin{equation}\label{eq:gravpoten}
   \nabla^2\Phi = \frac{16\pi G}{3}a^2\delta\rho_M + \frac{a^2}{6}\delta R(f_R),
\end{equation}
\begin{equation}\label{eq:deltafr}
   \nabla^2\delta f_R = -\frac{a^2}{3}[\delta R(f_R) + 8\pi G\delta\rho_M],
\end{equation}
where the gravitational potential is denoted by $\Phi$, the $f_R$ value is the scalaron, defined as $\frac{d f(R)}{dR}$, which is the extra scalar degree of freedom, $\delta R$ and $\delta\rho$ are the differences between the local values and the background values, denoted as $R - \bar{R}$ and $\rho -
\bar{\rho}$ respectively, with the barred quantities being the background values. Typically, the gravitational potential in general relativity (GR) depends only upon the matter distribution, $\delta\rho_M$, with $G$ remaining constant. However, with the addition of the scalar field, Newton's constant also becomes dependent upon $\rho_M$. In the limit of low, underdense regions, $G$ effectively becomes strengthened by a factor of $1/3$ as the second term in Eq. (\ref{eq:gravpoten}), $\delta R(f_R)$, vanishes. In the opposite limit, in high density regions, $\delta f_R$ in Eq. (\ref{eq:deltafr}) approaches zero, which sets $\delta R(f_R) = -8\pi G\delta\rho_M$. When Eq. (\ref{eq:gravpoten}) is re-evaluated with this, one sees that $\nabla^2\Phi$ now matches local GR. This is how the chameleon mechanism allows $f(R)$ gravity to pass solar system tests, crucial in making sure $f(R)$ remains feasible.

To follow the full non-linear evolution, Eqs (\ref{eq:gravpoten}) and (\ref{eq:deltafr}) cannot be solved analytically, so $N$-body simulations are required for analysis. For this work, we used the simulations described in \citet{Zhao:2011a}. The model of gravity assumed for the simulations was of the form $f(R) = \alpha R/(\beta R+\gamma$) where $\alpha = -m^2c_{1}$, $\beta = c_{2}$, and $\gamma = -m^2$, all determined by three underlying variables. Of the three, only one is predetermined, with $m^2 = H^2_{o} \Omega_{M}$. The other two variables, $c_1$ and $c_2$, are free parameters that determine both the expansion rate of the universe in the $f(R)$ model, given by the ratio $c_1/c_2$, as well as the rate of structure formation, which is proportional to $c_1/c^2_2$. The structure formation specifically depends upon the value of $|df/dR|$ at redshift zero, referred to as $|f_{R,0}|$. To ensure a valid comparison to \lcdm~cosmology, $c_{1}/c_{2}$ was set equal to $6\Omega_{L}/\Omega_{M}$, which provides the same expansion history as in \lcdm and an agreement of the value of $\sigma_8$ with GR at 
redshift 0. The values chosen for $|f_{R,0}|$, $10^{-4}$, $10^{-5}$, $10^{-6}$, pass current solar system tests. Hereafter these different models will be referred to as F4, F5, and F6, respectively. 

Six realizations were computed for each model, including GR (i.e., $|f_{R,0}| = 0$). Each simulation box contained $1024^3$ dark matter particles and had a cubic volume of 1.5 \gmpc~ per side. For this paper, we selected snapshots at scale factors of $a=0.7$ and $a=0.5$, corresponding to redshifts $z=0.43$ and $z=1.0$. The latter redshift represents the peak galaxy number density regime of the Euclid survey \citep{Euclid:2011} where the expected galaxy number density is roughly $\bar{n} = 1.6 \times 10^{-3}$ per cubic \hmpc~. At this redshift, the expected survey volume will exceed the simulation volume, which will help to decrease statistical errors further.
The redshift $z=0.43$ represents, under reasonable assumptions, the survey volumes for a spectroscopic ground-based mission such as DESI~\citep{DESI} where the galaxy number density is expected to be $\bar{n} = 2.6 \times 10^{-3}$ per cubic \hmpc~. 

We subsample the simulation dark matter particles to a mean density of $\bar{n} = 2 \times 10^{-3}$ per cubic \hmpc~ to provide a realistic tracer density in the void finding process, again to provide comparable results to future surveys. In addition, particle positions were perturbed according to their peculiar velocities to demonstrate the observable properties in redshift space, negating the need to correct for redshift-space distortions. It should be noted that although this paper aims to develop an intuition for the observational indicators of modified gravity, we choose to ignore the effects of galaxy bias, as \citet{Sutter2013a} demonstrated that void properties from catalogs compiled using a watershed based void finder are relatively insensitive to bias. For purposes of comparisons to other works, these results will be comparable to voids in dark matter halos and galaxies.

Voids were identified using the publicly available Void Identification and Examination (VIDE) toolkit \citep{Sutter2014e}\footnote{http://www.cosmicvoids.net}, which uses a substantially modified version of ZOBOV \citep{Neyrinck2008} to conduct a Voronoi tessellation of the particles and perform a watershed transform to group the Voronoi cells into a hierarchical tree of subvoids and voids \citep{Platen2007}. A final catalog of voids is then built using the criteria that the voids must be larger than the mean particle separation, and for our analysis we will only consider voids larger than 7~\hmpc~ that have central densities lower than 0.2 of the mean particle density $\bar n$.

% -----------------------------------------------------------------------------
% -----------------------------------------------------------------------------
\section{Results}
\label{sec:effects}

In our analysis of the simulations, we will examine in detail four properties: abundances, radial density profiles, radial velocity profiles, and ellipticities. Previous works have shown the susceptibility of these properties to changes in gravity (e.g.,~\citealt{Bos2012,Sutter2014d,Cai2014b}). This work focuses on the observational relevancy of each property, so we will present the analysis with emphasis on the strengths and weaknesses each property displays. For each property discussed below, the mean values correspond to the mean across all six realizations. We take the quoted variance to be either the cosmic variance or the intrinsic scatter, depending on which is larger, to be as pessimistic as possible. All properties were calculated using built-in functions in VIDE, with slight modifications made to accommodate handling multiple realizations.

\begin{figure}
  \centering
{\includegraphics[type=png,ext=.png,read=.png,width=\columnwidth]{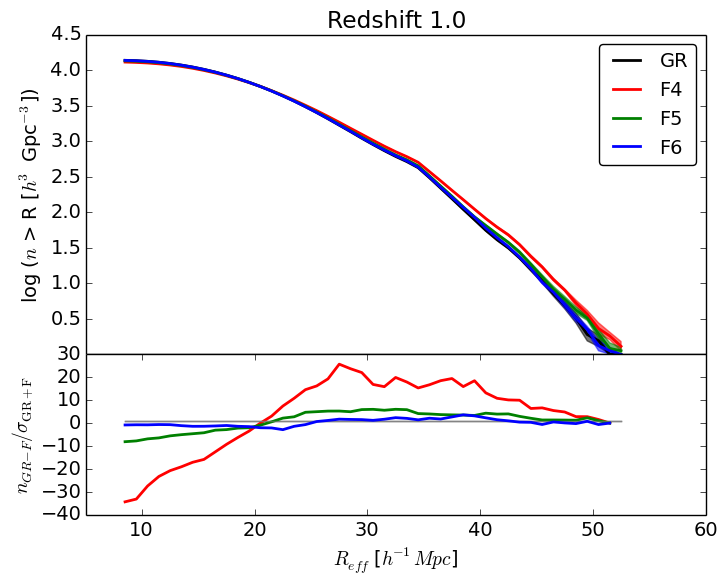}}
{\includegraphics[type=png,ext=.png,read=.png,width=\columnwidth]{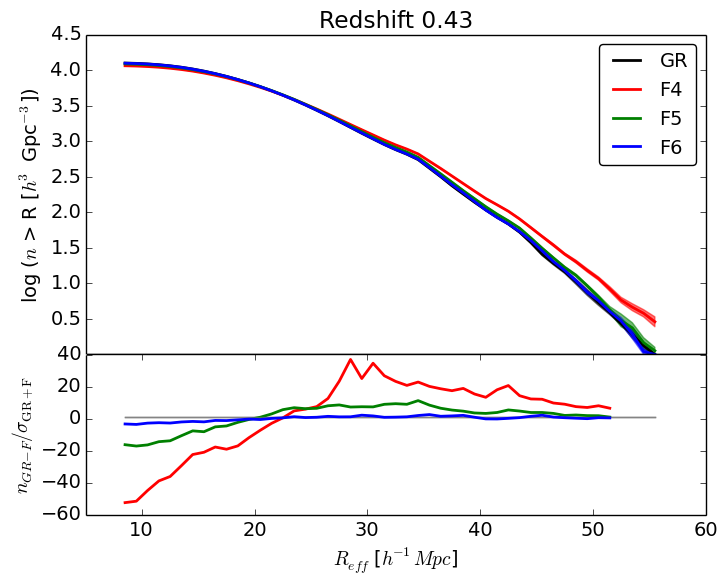}}
   \caption{Cumulative void number functions. The upper panels show the abundances for General Relativity (GR; black) and modified gravity models F4 (red), F5 (green), and F6 (blue) from realistically subsampled dark matter particle simulations. The solid lines are the  mean number functions of the six realizations, and the shaded regions are the 1$\sigma$ cosmic variances for each mean. The lower panels show the relative significance of each model compared to GR. Larger values of $|f_{R0}|$ cause the modified gravity to turn on at earlier ages, accelerating the evacuation of matter compared to GR, leading to fewer small voids and more large voids.
            }
\label{fig:numberfunc}
\end{figure}

Figure~\ref{fig:numberfunc} shows the cumulative number function from \lcdm~ and the modified gravity simulations at redshifts $1.0$ and $0.43$. Along the x axis is the effective void radius (the radius of a sphere with the same volume 
as the void). On the top portion of the plot is the log of the number of voids larger than a given effective radius per cubic \gmpc, and in the lower portion is the relative significance of the models F4, F5, and F6 compared to GR. The uncertainty plotted in the top portion is the square root of the cosmic variance. For the bottom plot, the difference between the number of voids in the GR model and in the $f(R)$ models was divided by the square root of the sum of variances from the two models. This provides a weighted indication of where the differences provide the most statistically significant signal. 

We can see that F4 clearly contains larger voids than in the \lcdm~simulation at both redshifts. Even at the higher redshift, in the roughly 40 \hmpc~ radius range, there are still enough voids to provide significant statistical power in distinguishing between the F4 model and GR, clearly indicated by the relative significance plots. 
It becomes difficult to see any clear differences between the weaker coupling strengths and GR, especially at redshift $1.0$. However, due to the large numbers of voids being examined, the errors on the abundances for F5 become small enough to separate from that model from GR, most notably in two separate regimes, the small voids on the order of ~$10$ \hmpc~ and the medium-large voids on the order of $~30$ \hmpc. While the differences can be detected at both redshifts, at the lower redshift, the statistical power, especially in smaller voids, doubles by nearly a factor of two. 

The obvious tilt in the number function reveals the effect of the modified gravity on the structure of voids: the smaller voids in GR have been emptied out and the void walls have begun to thin, allowing the watershed algorithm to merge them together, as seen in the coupled DM-DE analysis of \citet{Sutter2014c}. This produced few small voids and more large voids.
However, despite the $\sim 20\%$ differences in the number of large voids, 
the peak relative significance occurs for small and medium-scale voids, 
where the increased statistical significance overcomes the relatively 
smaller absolute differences.
Similarly, even though the modified gravity models produce larger 
absolute differences at lower redshifts, there are few overall voids, 
so the relative significance remains largely unchanged.

The evolution of these differences align with what one would reasonably expect to see from the $f(R)$ models. At higher redshift, the voids have not had time to empty out. Because the modified gravity from the scalar field in $f(R)$ is dependent upon the local density, until voids empty out enough, the modified gravity will remain screened, making the $f(R)$ models appear identical to GR. Simultaneously, as time progresses and observations move to lower redshifts, they see an older universe, one that has given the modified gravity more time to act upon the voids, expanding them more rapidly than normal GR would, resulting in a higher overall number of large voids. The ordering that is prominently displayed in the lower redshift plots stems from the differing strengths of the modified gravity, with the strongest force, F4, producing the greatest number of large voids. 

To provide an initial estimate of the ability of these voids to constrain the value of $|f_{R,0}|$, we perform a simple Fisher forecast by constructing a numerical derivative of the abundances as a function of parameter strength:
\begin{equation}\label{eq:fisher}
	\Delta f = \sum\limits_{i=1}^N \frac{ (n_{ {\rm F4},i}-n_{{ \rm GR}, i } )^2}{10^{-4}\times \sigma^2_{F4,i}},
\end{equation}
where $\Delta f$ is the resulting forecasted upper limit, $n_{ {\rm F4}, i}$ and $n_{ {\rm GR}, i}$ denote the void number density in radial bin $i$ for the F4 and GR simulations, respectively, $N$ is the total number of radial bins, $\sigma^2_{F4}$ is the variance in a bin, and $10^{-4}$ is the difference in the $|f_{R0}|$ parameter value between the F4 and GR model. Using this prescription, we find that we can place an upper limit of $5.82\times 10^{-5}$ for voids at redshift $1.0$ and $4.78\times 10^{-5}$ at redshift $0.43$ on the value of $|f_{R,0}|$. As one might expect, with the greater differences between models at more recent times, we are able to more tightly constrain the value. With upcoming surveys, if more redshift epochs are available for analysis, the combined information may even be able to rule out F5.

\begin{figure*}
   \centering
{\includegraphics[type=png,ext=.png,read=.png,width=0.48\textwidth]{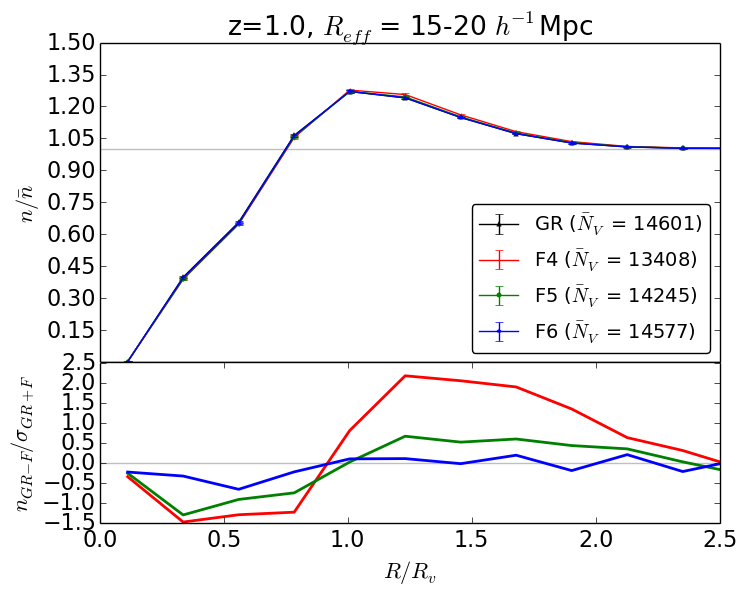}}
{\includegraphics[type=png,ext=.png,read=.png,width=0.48\textwidth]{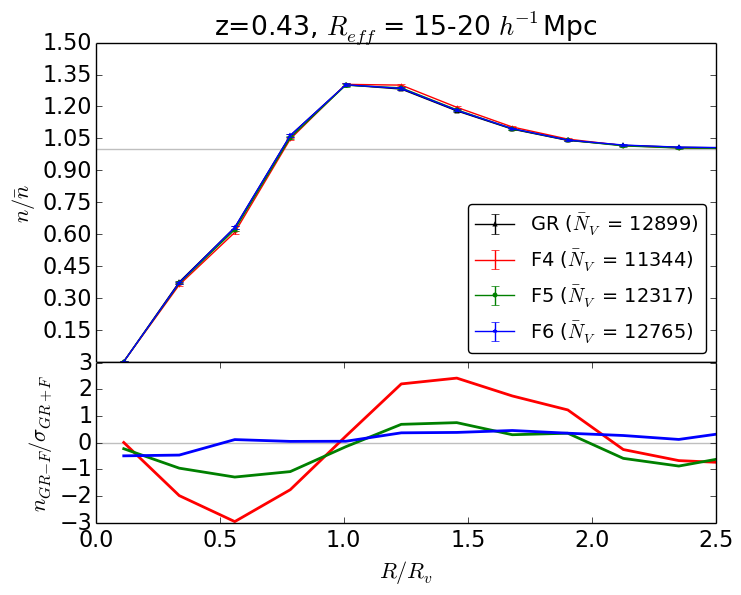}}
{\includegraphics[type=png,ext=.png,read=.png,width=0.48\textwidth]{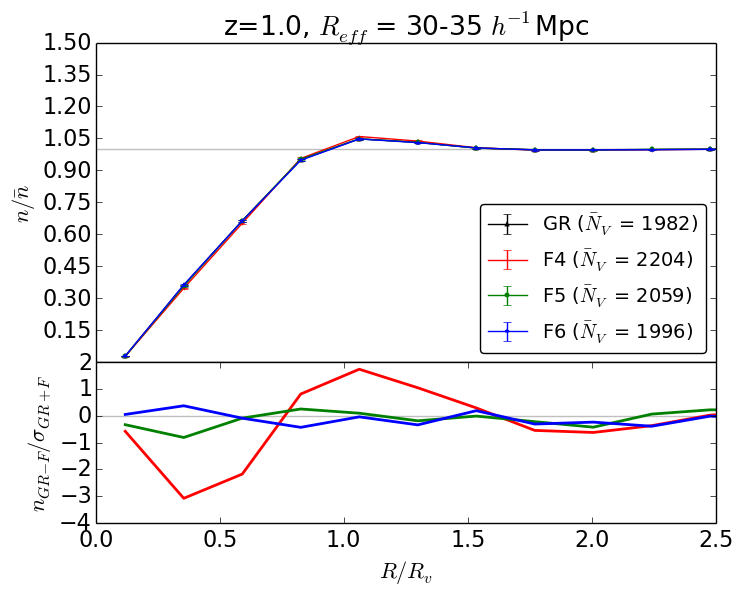}}
{\includegraphics[type=png,ext=.png,read=.png,width=0.48\textwidth]{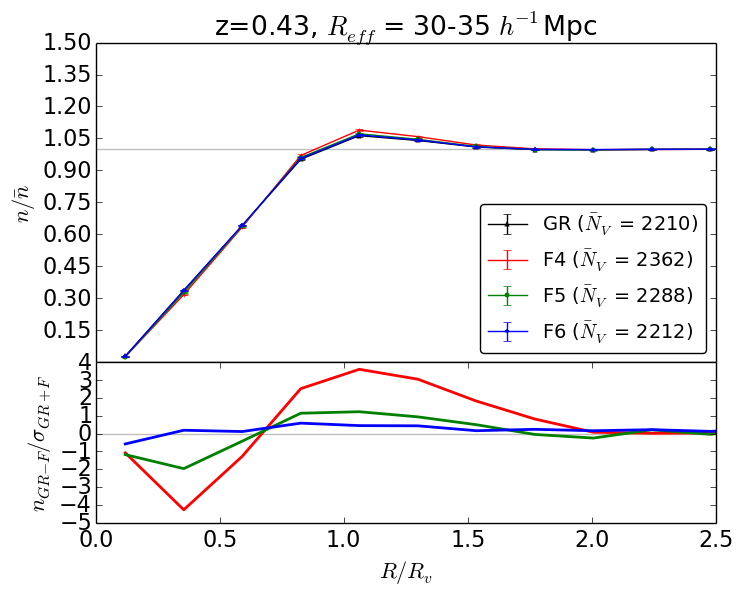}}
    \caption{Mean radial density profiles of stacked voids with 1$\sigma$ uncertainties (upper panels) and relative significance of the profiles compared to GR (lower panels). The legends indicate the cosmological model and mean number of voids $N_v$ used to calculate the profiles. The thin gray lines represent mean density in the upper panels and the GR relatively profile in the lower panels. The density is plotted against the relative radius, where $R_v$ is the median void radius in each stack. The accelerated evacuation under modified gravity generates a slightly greater build up in the compensation region, more noticeable at a lower redshifts.
            }
\label{fig:1d_profile}
\end{figure*}

In Figure~\ref{fig:1d_profile} we show the mean one-dimensional radial  density profiles for two different radial bins, 15-20 \hmpc~ and 25-30 \hmpc~  (representing, respectively, over- and under-compensated voids) for the six realizations. On the bottom portion of each plot, we show the relative significance between each model. In the legend, beside each model type is the total number of voids used to calculate the mean profiles. Within each realization, we computed the profiles by selecting voids within a given radial range and aligning their volume-weighted centers. With a sufficient number of voids, the average shape of a void approaches a sphere, so the radial densities were measured by taking the number of particles within thin spherical shells. We normalized these densities to the mean number density of the realization and plotted them against the effective radius, $R/R_v$, where $R_v$ is the median void size in each stack. 

It is important to note that error bars were plotted for the normalized plots, but due to the high number of voids used to calculate the profiles, the error bars are not visible. 

Upon first examination, one does not notice any drastic differences in the normalized profiles between the models in either redshift. 
However, looking at the relative significance plots below the normalized profiles, one sees that the differences in the compensation regions are in fact significant by up to 3$\sigma$. 
The differences are most apparent in the compensation shells surrounding the 
voids, where changes to void histories lead to differences in the pile-up 
of matter surrounding them.
The interiors of the voids are correspondingly emptier in the modified gravity models, with the switch between relatively over and under dense (compared to GR) appearing mid-wall. At higher redshifts, this switch occurs at higher radii for stronger 
gravity models, but by lower redshifts the switch occurs at nearly 
the same radius, since further void evolution is constrained by surrounding 
structures.

By $\sim 2 R_{\rm eff}$ the differences disappear.
Similarly, at the void centers the uncertainties are so large that 
the relative significance approaches zero.

Similar to the abundances, there is an ordering of the models with the weaker modified gravity models being closer to GR, as expected. 
We see here the impacts of modified gravity on void evolution. The enhanced gravity acts on the particles within the voids, accelerating them to the edges of the voids faster than GR would, causing the interiors of the voids to be less dense compared to GR while creating a build up of particles in the walls of the voids, leading to the denser overcompensated regions. As the voids are given more time to evolve and for the modified gravity to act, these differences become even larger, as one can discern from the transition of the plots from redshift $1.0$ to redshift $0.43$. However, one will notice that the weakest modified gravity model F6 does not have a noticeable trend in the significance plot, instead fluctuating across the GR baseline value. The holds for all four profiles examined, with the relative significance never rising above 1$\sigma$. Even F5 struggles to differentiate itself, with its significance values only just rising above 1$\sigma$ for redshift $1.0$. Only when the voids have had more time to evolve to redshift $0.43$ are there any consistent differences for F5.

Despite the lower significance of these differences, void density profiles 
may be an appealing target for future surveys, 
since they can be accessed in real space 
in a parameter-free way~\citep{Pisani2013} and are not affected 
by survey masks~\citep{Sutter2013c}.

\begin{figure*}
   \centering
{\includegraphics[type=png,ext=.png,read=.png,width=0.48\textwidth]{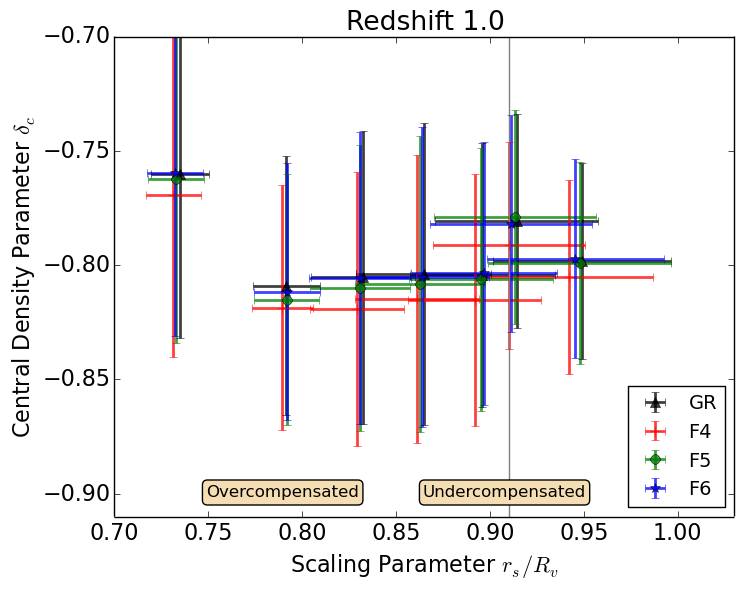}}
{\includegraphics[type=png,ext=.png,read=.png,width=0.48\textwidth]{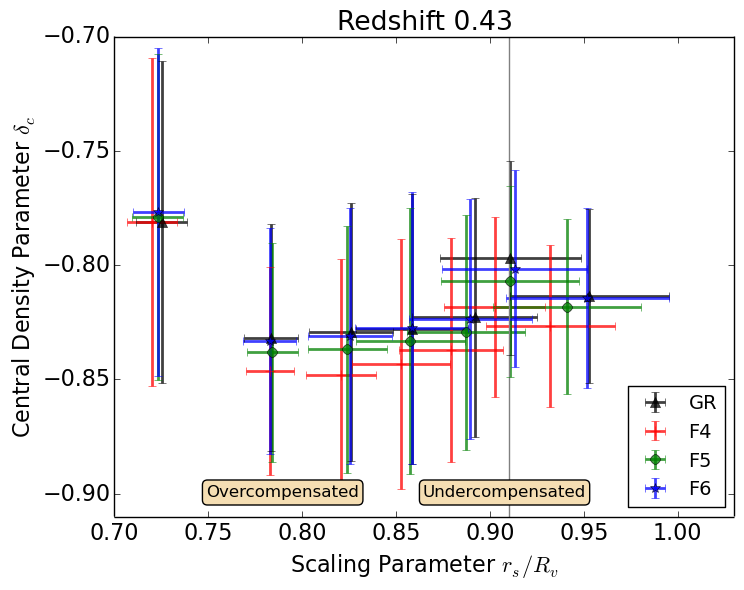}}
    \caption{Best-fit values and $1\sigma$ uncertainties for the stacks of voids using the HSW profile~\citep{Hamaus2014}. The thin grey line indicates the analytically-derived compensation scale. From left to right, the void stacks are of radius 10-15, 15-20, 20-25, 25-30, 30-35, 35-40, and 40-45 \hmpc.
The fits show a small but systematic trend: modified gravity acts to empty out the 
voids at all scales.
            }
\label{fig:hsw_profile}
\end{figure*}

One other potential route to extract more information from the radial density profiles is to fit them using the profile model proposed in \citet{Hamaus2014}, hereafter referred as the HSW profile. This model provides a universal void profile that requires only two parameters, the central density of the voids $\delta_{c}$ and the scaling radius $r_{s}$. Fitting to this profile allows us to neatly summarize any systematic differences between modified gravity models and GR. 
We split the voids into bins by radius, going from $10-15$ \hmpc~ in steps of 5 \hmpc~ up to $40-45$ \hmpc. The results of the fit can be seen in Figure~\ref{fig:hsw_profile}. In the plots the thin gray line indicates where the voids transfer from being overcompensated to undercompensated.

While individual stacks do not provide statistically significant differences,
we may discern a systematic trend: modified gravity models produce voids 
with lower values of both $\delta_c$ and $r_s$, with F4 producing the 
largest differences. The fits to the HSW capture the quintessential 
differences between the models and GR: modified gravity produces emptier, 
more steeply-walled voids.

To quantify the significance of these differences, we found the weighted distance between each fit using the prescription
\begin{equation}\label{eq:hswdist}
d = \sqrt{\frac{(x_{GR}-x_{F})^2}{\sigma^2_{x,GR}+\sigma^2_{x,F}}+\frac{(y_{GR}-y_{F})^2}{\sigma^2_{y,GR}+\sigma^2_{y,F}}},
\end{equation}
where x refers to the $r_{s}/R_{v}$ value, y refers to the $\delta_{c}$ value, F refers to the modified gravity model (as this distance was calculated for the three models, F4, F5, and F6), and $\sigma^2$ refers to the variance in the fit for the respective values. This acts as an analog to the relative significance plots from the radial densities and the abundances. Using this prescription, we find that the total significances for F4 are 1.0 and 1.8, for F5 are 0.4 and 0.7, and for F6 are 0.3 and 0.3, for redshifts 1.0 and 0.43 respectively. While it comes as no surprise that the F6 model struggles to separate itself from GR, even under this severe compression of the data the F5 and F4 models are only 
mildly distinguishable from GR, showing a significance lower than the radial density profiles displayed at their peaks.

\begin{figure*}
   \centering
{\includegraphics[type=png,ext=.png,read=.png,width=0.48\textwidth]{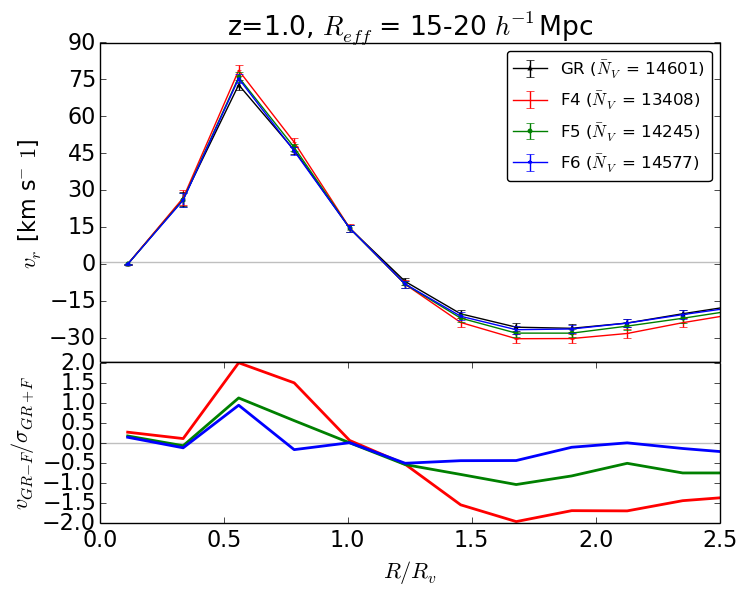}}
{\includegraphics[type=png,ext=.png,read=.png,width=0.48\textwidth]{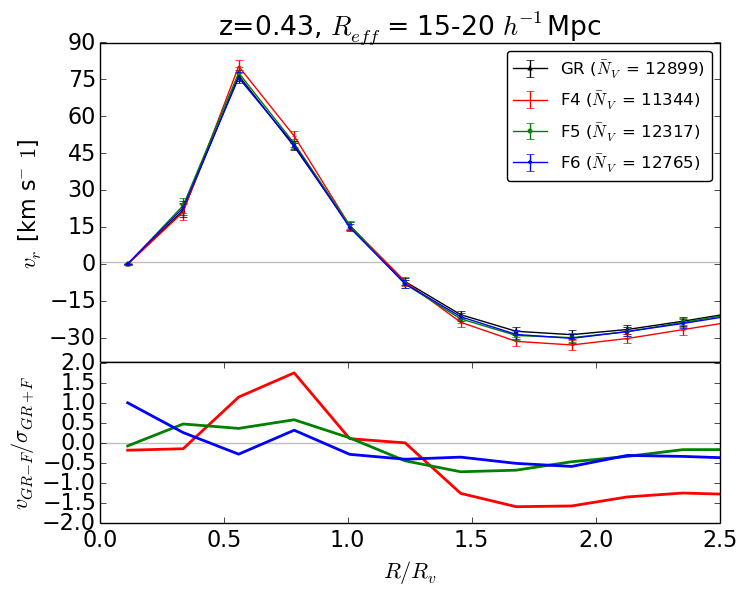}}
{\includegraphics[type=png,ext=.png,read=.png,width=0.48\textwidth]{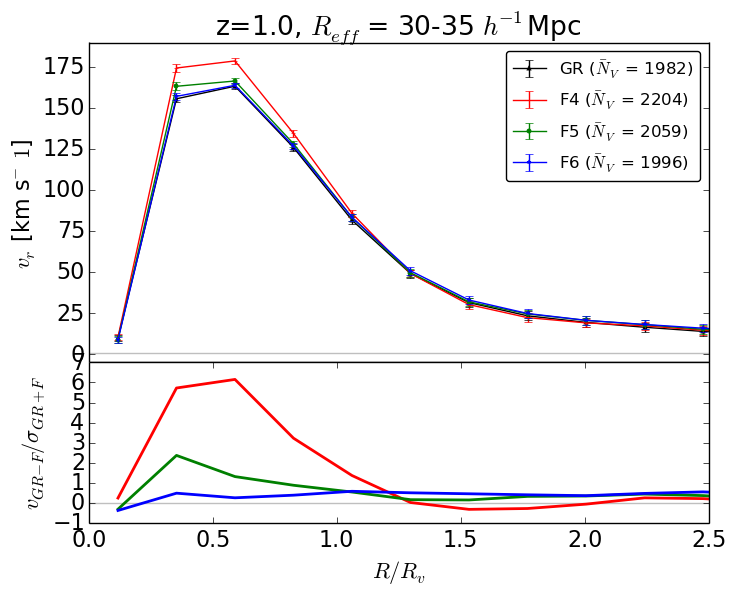}}
{\includegraphics[type=png,ext=.png,read=.png,width=0.48\textwidth]{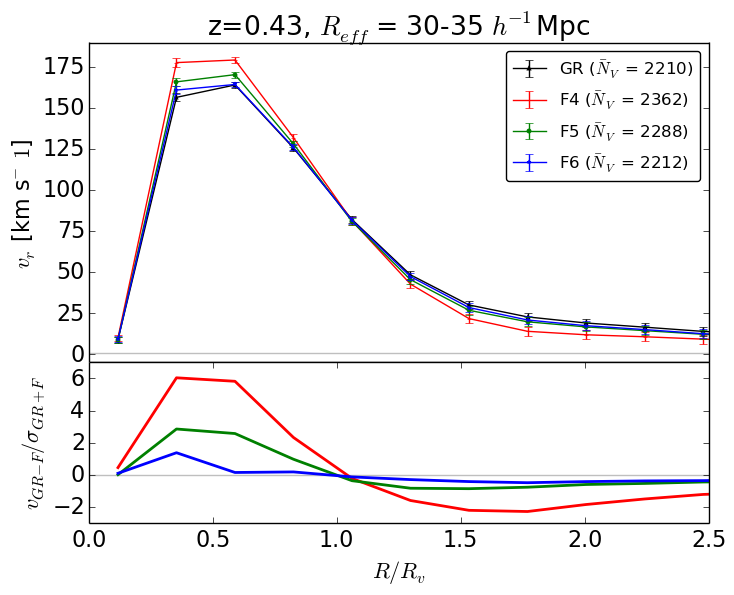}}
    \caption{Mean radial velocity profiles of stacked voids with 1$\sigma$ uncertainties (upper panels) and relative significance of the profiles compared to GR (lower panels). The legend in each plot indicates the mean number of voids $N_v$ used in total to calculate profiles. The radial velocity is plotted against the relative radius, where $R_v$ is the median void radius in each stack. Note for larger voids, the models are more easily distinguished, and there is minimal diminishing of the significance at higher redshifts.
While the absolute differences between the modified gravity models and GR 
is larger than in the density profiles, the velocities carry greater uncertainties.
		}
\label{fig:vel_profile}
\end{figure*}

In Figure~\ref{fig:vel_profile} we plot the radial velocities profiles for the same radius stacks as the density profiles. As before, the top panels show the profiles with 1$\sigma$ uncertainties and the bottom plots show the relative statistical significances. Positive values indicate outflows and negative values 
indicate inflow relative to the void center.
As expected, we see that the larger voids have much faster outflow velocities, 
by up to a factor of two. This is because the larger voids are not surrounded 
by overdense shells, and thus do not have their expansion constrained by 
any surrounding matter. For a similar reasons the velocities never drop below 
zero for the largest voids, indicating that they will continue expanding, 
unlike their smaller, overcompensated counterparts.

The differences between the modified gravity models and GR are visually 
apparent, in contrast with the density profiles. Under the influence 
of modified gravity, the peak outflow velocities at late times 
increase by up to 
$\sim 20\%$ in the F4 model. The corresponds to an reduced 
outflow outside the void for larger stacks and an enhanced inflow 
for smaller stacks. Since modified gravity directly affects acceleration, 
it is not surprisingly that velocities will be impacted more than densities.
However, velocities also have considerably more scatter, so despite these
strong differences, the relative significance remains similar to what we observed earlier in the radial density profiles.

The velocity profiles offer one unique advantage compared to the other void properties previously examined: a strong difference that persists at higher 
redshifts. Indeed, the void interiors have more significant differences 
at higher redshifts than at lower ones, since the growth of the voids has 
not yet been strongly affected by their surroundings (i.e., either by 
the build-up of an overdense shell or by running into adjacent voids). 
The changes already present at high redshift in the velocity profiles will 
not manifest themselves in the density profiles or abundances until 
later times.

Interestingly, the distinguishing characteristic in signal between categories of voids appears to be the size of the voids instead of the redshift. One will note that for the smaller voids the significance even for F4 remains roughly constant at both redshifts at around 2$\sigma$. The larger voids appear to have a much higher significance, closer to 6$\sigma$, for the outflow velocities instead the voids themselves. This significance remains even at the higher redshift, unlike the radial density profiles.

Velocities then potentially offer better leverage in the higher redshift surveys than density profiles, but it still cannot begin to match the abundances in terms of overall significance.

\begin{figure*}
  \centering
{\includegraphics[type=png,ext=.png,read=.png,width=0.48\textwidth]{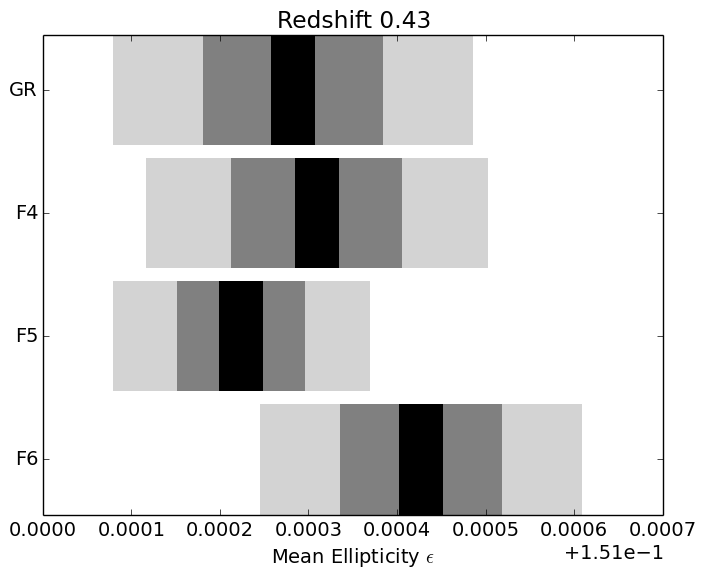}}
{\includegraphics[type=png,ext=.png,read=.png,width=0.48\textwidth]{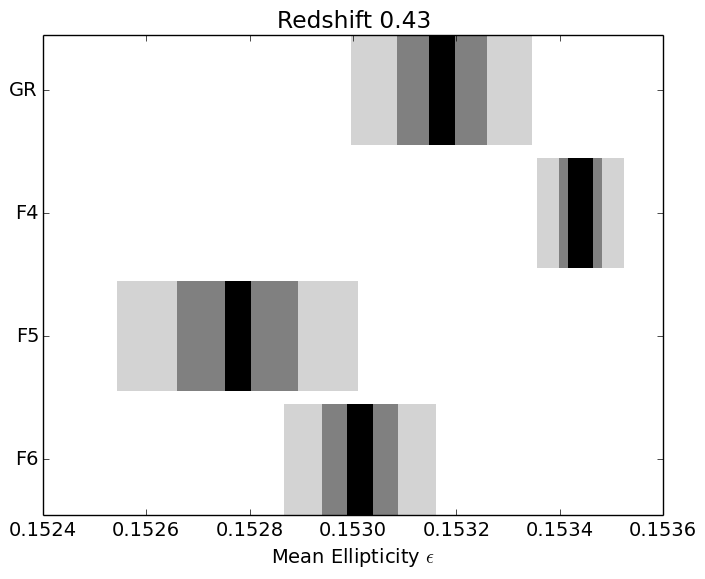}}
    \caption{Mean ellipticity (black) with 1$\sigma$ (dark grey) and 2$\sigma$ errors on the mean for each model. All voids are used to calculate the ellipticity distributions.
	}
\label{fig:ellip}
\end{figure*}

The final void property we examine is the ellipticity. In Figure~\ref{fig:ellip}, histograms show the mean void ellipticity of each model, marked by the black line, and the $1 \sigma$ and $2 \sigma$ errors on the mean, marked by the darker gray and lighter gray areas respectively. 
To calculate the ellipticity we use the 
inertia tensor to compute eigenvalues and
eigenvectors and form:
\begin{equation}
  \epsilon = 1- \left( \frac{J_1}{J_3}\right)^{1/4},
\label{eq:ellip}
\end{equation}
where $J_1$ and $J_3$ are the smallest and largest eigenvalues
of the inertia tensor, respectively.

Unlike the properties studied above, there is no clear ordering from the strongest coupling strengths to the weakest coupling strengths. 
The notable outlier at both redshifts is the strongest coupling strength, F4, which is a few $\sigma$ removed from the means of the other models. However, aside from this outlier, there is no readily discernible pattern to be seen. For the higher redshift, none of the mean ellipticities appear to be statistically significant from each other, suggesting an overall similar shape to the voids at this point in their evolution. At a later time the voids have evolved further to begin to have a set shape in the universal environment, but there is no apparent pattern in how the mean ellipticities are chosen for each model as the weakest modified gravity model, F6, has a larger average ellipticity than F5, while GR is stronger than both of them.

Since modified gravity distorts the size distribution of the void population,
we can understand these differences by considering the ellipticity of voids as a function of size. Smaller voids are highly elliptical, since they are bounded 
by irregularly-shaped walls, medium voids are more isotropic, and finally 
the largest voids are once again highly elliptical, since their volume-filling 
nature prevents them from growing uniformly in all directions~\citep{Sutter2013c}. The F4 model produces a significantly enhanced population of larger voids, 
thereby skewing the mean ellipticity. The fact that the F5 and F6 models do not appear significantly different from GR indicates that they have not distorted the void population to such a degree as to influence the ellipticities. 

Similar to most of the previous properties, the differences between the models become more distinct at lower redshifts. This can be caused by several factors. One potential explanation is that structure may grow faster overall in modified gravity models, enhancing differences in ellipticities over time. Another explanation is the relationship between void evolution and the screening mechanism. Although modified gravity only depends on the depth of the local Newtonian potential, rather than its gradient, the potential within a void is not uniform, and thus the effects of modified gravity can be anisotropic. As the underdensities begin to form, not every direction will empty out uniformly, leading to an anisotropic potential. Particles near the ends of the ellipse, where the potential is lowest, will become unscreened soonest, exaggerating the initial axis of expansion. However, this only applies for smaller voids as their small size will highlight any imbalances in the eigenvalues. As they expand, the ratio between the eigenvalues will begin to shrink as the modified gravity acts upon the other axes of the void. However, small voids comprise enough of the total void population to still make this a potentially observable effect.

% -----------------------------------------------------------------------------
% -----------------------------------------------------------------------------
\section{Conclusions}
\label{sec:conclusions}

We have performed an initial assessment of the ability of upcoming galaxy surveys to distinguish models of modified gravity from general relativity using voids.
Modifications to gravity that can be used to explain the accelerated expansion of the universe also manifest themselves as different void populations compared to general relativity. We have examined simulations including modified gravity that are designed to have identical large-scale clustering statistics as general relativity, but despite this the voids in modified gravity appear larger with steeper density profiles.  The HSW profile fits show a systematic impact on the profiles: voids are emptier and steeper under modified gravity. This is a direct consequence of the additional acceleration present in unscreened underdense environments. As the simulations evolve, the modified gravity has longer to operate and enhances these differences. At lower redshifts, we observe that the differences in the abundances becomes more pronounced, with the creation of even larger voids and the merging of smaller voids into larger ones.  

In the context of upcoming galaxy redshift surveys, these differences are potentially detectable.
A simple Fisher forecast for $|f_{R,0}|$ using the void abundances places an upper detection limit of about $5\times 10^{-5}$, which means that upcoming surveys may be able to rule out the F5 model. The density profiles for the low redshift survey volumes provide additional constraints, although with the low tracer density the information available only shows a roughly 2$\sigma$ difference. While this does indicate a potentially detectable difference, its impact is not as large as initially suspected. The radial velocity profiles provide a large constraint, showing up to 6$\sigma$ differences in both low and high redshift survey volumes. In particular with a low tracer density, the distribution of void sizes shifts to produce larger voids, in turn providing more voids that possess high outflow velocities. Of course, while the initial assessment is encouraging, radial velocity measurements possess significantly more difficult challenges and errors than measuring the positions of tracer galaxies, as the uncertainty in velocity increases with redshift and that due to the isotropic nature of voids, only roughly one in ten galaxies will be oriented in such a way as to allow for a measurement of its radial velocity. However, with the large volume of Euclid, it may be feasible to collect enough outflow velocities of voids to offset this difference in data points. Overall, the statistical significance of these profile differences falls nearly an order of magnitude below the significance found in the void abundances, which while susceptible to changes in the tracer density still appear to provide larger relative significances even when the tracer density varies by a factor of 2, a positive sign for the ability of ground based surveys examining nearer sections of the universe to detect divergences from expected GR void population statistics. These divergences will benefit from additional testing through techniques being developed for modeling dark matter profiles with high precision which will help to more cleanly distinguish between the models~\citep{Leclercq2015}.

\citet{Cai2014b} have recently performed a similar analysis using a spherical-underdensity void finder. While they do observe differences in the population of small voids, our watershed technique reveals significantly different small void populations. Indeed, most of our statistical significance derives from the small-void end of the abundance functions. Watershed-based void finding also appears more robust: the differences in void properties as a function of coupling strength maintain the same ordering at different sparsity levels, unlike the ~\citet{Cai2014b} analysis. We also see more significant differences in the density and velocity profiles. Finally, watershed voids are much less sensitive to galaxy bias, as shown by~\citet{Sutter2013a}. 

We have performed only an initial assessment, although our study includes several realistic aspects, such as sparsity and peculiar velocities. A more complete analysis would model lightcone and masking effects, and will be included in future work.
It should be noted that future galaxy surveys will capture \emph{at least} as many voids as we have studied here. At redshift $1.0$ our simulation underestimates the volume --- and thus the number of voids captured --- by Euclid and WFIRST.
We have seen in our analysis that for detecting modified gravity there is a trade-off: higher redshifts give access to smaller statistical uncertainties, since there are more voids overall, but the modified gravity effects have not had time to largely impact void properties. It appears that these competing effects balance each other, and a space-based high-redshift survey delivers roughly comparable constraints as a ground-based low-redshift survey.
Thus voids appear as a promising avenue for exploring and constraining modified gravity models that are inaccessible to traditional probes.

%-------------------------------------------------------------------------------
%-------------------------------------------------------------------------------
\section*{Acknowledgments}

The authors thank Nelson Padilla, David Weinberg, Nico Hamaus, and Alice Pisani  for useful discussion. PZ thanks OSU CCAPP 
for summer research support. PMS is supported by the INFN IS PD51 "Indark". 

\footnotesize{
  \bibliographystyle{mn2e}
  \bibliography{frindicator}
}

\end{document}